\documentclass[
reprint,
amsmath, 
aps,
onecolumn, 
superscriptaddress,
]{revtex4-1}

\usepackage[colorlinks=true,linkcolor=red]{hyperref}%

\usepackage{graphicx}
\usepackage[english]{babel}
\usepackage[utf8]{inputenc}
\usepackage[colorinlistoftodos]{todonotes}

\def\beq{\begin{equation}}
\def\eeq{\end{equation}}

\def\bs{\begin{split}}
\def\es{\end{split}}

\def\bw{\begin{widetext}}
\def\ew{\end{widetext}}

\def\bep{\begin{pmatrix}}
\def\eep{\end{pmatrix}}

\def\bea{\begin{eqnarray}}
\def\eea{\end{eqnarray}}

\def\beq{\begin{equation}}
\def\eeq{\end{equation}}
\def\bea{\begin{eqnarray}}
\def\eea{\end{eqnarray}}
\def\brcl{\begin{array}{rcl}}
\def\bccl{\begin{array}{ccl}}
\def\blcl{\begin{array}{lcl}}
\def\err{\end{array}}

\def\fatk{{\mathbf{k}}}

\def\ef{{$\varepsilon_{\rm{F}}$}}
\def\ef0{{\varepsilon_{\rm{F}}}}

\def\enk{{\varepsilon_{i\mathbf{k}}}}

\begin{document}

\title{Fermi softness: a local perspective on surface reactivity}

\author{Bing Huang}
\affiliation{Institute of Physical Chemistry and National Center for Computational Design and Discovery of Novel Materials (MARVEL),
Department of Chemistry, University of Basel, Klingelbergstrasse 80, 4056 Basel, Switzerland}

\author{Lin Zhuang}
\affiliation{College of Chemistry and Molecular Sciences, Wuhan University, 430072, Wuhan, China}






\begin{abstract}
Understanding how electronic structure determines the reactivity of solid surface, is a central topic of modern surface science.
This is mostly commonly done through some intermediate quantity termed descriptor.
However, such descriptors are very scarce for solid surface compared to for molecule, likely due to their significantly more complex electronic structure (cf. molecules).
Here we elaborate on the theory of a concept dubbed ``Fermi softness'', which distinguishes itself by enabling prediction of surface reactivity with spatial as well as atomic resolution. 
Other pertinent concepts and descriptors are also mentioned so as to make the treatment comprehensive.
\end{abstract}


\maketitle
\tableofcontents

\section{Introduction}

Since the very early days of theoretical chemical research, people have been enthusiastic in understanding how structure, be it geometrical or electronic,
translates into the behavior of a chemical species,
being either inherent property of the species \emph{per se} or its interaction strength with the other incoming species.
Of particular interest is the latter problem concerning the interaction between two reactive species,
which also constitutes part of the grand challenge in modern pursuit of conceptual understanding of the mysterious nature of chemical bonds~\cite{pauling_nature_1960,ruedenberg_physical_1962,sousa_are_2017,chemplitude,shaik_charge_shift_2009,shaik_charge_shift_2020,levine_energy_2017,levine_quantifying_2017,levine_clarifying_2020}.


Admittedly it would be very complicated (and likely demanding) a problem to accurately examine the interaction process in detail, which typically relies on expensive quantum chemical calculations (or machine learning techniques~\cite{huang2020chemrev}), or through sophisticated experimental setups.
A vast swathe of studies have been conducted in this direction and a lot of insights have been obtained thereafter.

Yet a different approach is to distill the most vital information from a single (or very few) static calculation(s) of the standalone species involved in interaction, or some simple experimental quantities (e.g., ionization potential).
In the second approach, we are most interested about the general trend of interaction strengths and some derived simple physical quantity across different systems (even better if the established reactivity model offers quantitatively correct predictions), with only one constitute varying (e.g., A interacts with a series of B's, where B's are not very different but share some common feature in geometrical/electronic structure).

As can be seen, the second approach emphasizes the conceptual aspect of chemical interaction,
and offers insight into chemistry in spite of the fact that its application is limited (cf. the first \emph{ab initio} approach) and may only make qualitatively correct predictions.

Early such practises comprise almost exclusively simple linear (or low dimensional) correlations, with the underlying physics largely unknown, and one of the most notable of which being the so-called free energy relationships, e.g., Bell-Evans-Polanyi (BEP) principle~\cite{BEP_1,BEP_2,van_santen_BEP_TMsurf_2010}.
With the advent of quantum theory, in particular density functional theory,
relevant stories have been completely rewritten:
most of the previous empirically proposed chemical concepts such as
electronegativity~\cite{mulliken_new_1934}, hardness/softness~\cite{parr_absolute_1983,yang_hardness_1985,nguyen_local_2003}, etc.
could be rationalized.
This specific subject, dubbed ``conceptual DFT'' (CDFT for short), pioneered by Parr, Yang and co-workers~\cite{parr_cdft_book_1994}, has been extensively developed ever since and championed by many, including Geerlings,  De  Proft,  Ayers and co-workers~\cite{ayers_variational_2000,geerlings_conceptual_2003,geerlings_conceptual_2014}.

Within the first-principle view, the distillation can be viewed as an irreversible process of abstracting essential information from the system wavefunction $\Psi$ (obtained through solving Schrodinger equation of standalone entities before interaction).
The distilled quantity is often termed ``descriptor'' and in principle to obtain which it should not be too involved.
Note that how the distillation is carried out depends on the nature of the problem at hand,
and hardly there exist a universal descriptor that controls every behavior of a system interacting with the other.

For molecules, profuse knowledge have been accumulated through decades of theoretical research. For a comprehensive review, the reader is referred to Greeling's several review papers~\cite{geerlings_conceptual_2003,geerlings_conceptual_2014}. 
When it comes to solid and solid surface, however, well-established reactivity descriptor for molecules cease to work within their native form.
And often, extension of the same quantity to solid/surface is highly non-trivial or not feasible at all, due primarily to the significant difference between the electronic structure of molecule and solid/surface. 
More specifically, solid/surface possess much more complicated electronic structure compared to that of molecule, i.e.,
continuous band structure for the former vs. discrete energy levels for the latter.
Another unfavorable consequence is the new emerging interaction patterns between empty states of adsorbate and that of surface, further plagues the problem.
For a comprehensive review on these conceptual understanding, the reader is referred to the pioneering work done by Hoffman~\cite{hoffmann_how_1987,hoffmann_chemical_1988}.
Therefore, discovery of new working descriptors for surface is usually not a smooth process
and it is not trivial to derive such descriptor purely from fundamental theories.
Nevertheless, a multitude of descriptors regarding solid surface, have been proposed, including 
early experimental discovery of the correlation between the catalytic reaction rate and the work function for some solid surface by Vayenas et al.~\cite{vayenas_dependence_1990},
and the more concerned theoretical works, among which, notable ones consist of
Yang et al.'s work of extending the concept of softness/hardness from molecule to metallic surface~\cite{yang_hardness_1985},
Wilke et al.'s local isoelectronic reactivity of solid surfaces~\cite{wilke_local_1996},
and recently
Calle-Vallejo and co-workers' generalized coordination number~\cite{callevallejo_fast_2014,calle-vallejo_finding_2015,calle-vallejo_introducing_2015},
Ma et al.'s orbital-wise coordination number for predicting adsorption properties of metal nanocatalysts~\cite{ma_orbitalwise_2017}, etc.


Generally speaking, reactivity theory mainly deal with early stages of chemical reaction.
This is especially true for interactions involving two (or more) molecules, 
as atoms would rearrange themselves to an extent such that
information about the reactant would be lost at the final stage of reaction.
Put it another way, reactivity theory is likely to fail for such ``complicated'' interactions.
In contrast, solid surfaces or nanoparticles are relatively open, and its interaction with simple molecular species (e.g., an free O atom or a small radical OH) would barely cause any significant rearrangement of the constituting atoms.
Therefore, the knowledge obtained (through quantum chemical calculation or experiment) for reactants, e.g., relative magnitude of some reactivity index, may be preserved for products and translate well into the trend of interaction strength among different systems.
This openness may be responsible for the success of a wide range of surface reactivity indices and we will come back to this point later.


Hereafter, unless otherwise stated, the systems we deal with throughout this text are all metallic (i.e., no gap in band structure),
as this type of system is most studied and represents the most common catalyst in practise.

\subsection{d-band center model}
Before stepping into the formal introduction of Fermi softness, it is deemed necessary to
briefly discuss the $d$-band center model (or simply $d$-band model) for transition metal surface developed by Hammer, N{\o}rskov and co-workers~\cite{hammer_electronic_1995,hammer_why_1995}, 
as it is highly relevant to our study, and will also be used as a reference for model comparison, shortly.

\begin{figure}[h!] 
\includegraphics[scale=0.6]{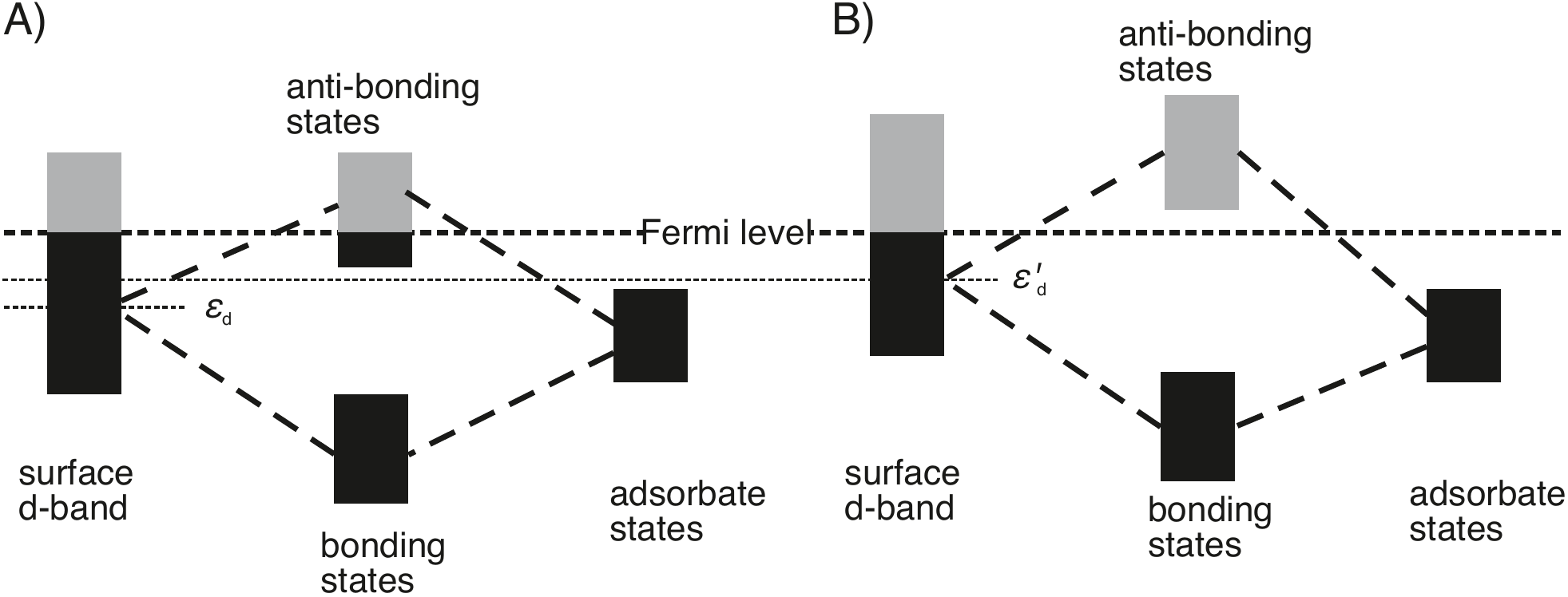}
\caption{Schematic illustration of stabilization of total single electron energies upon shifting from $\varepsilon_d$ (A) to a higher value ($\varepsilon_d'$, B) of $d$-band center of transitional metal surface when interacting with an adsorbate.}
\label{fig:d-band-model}
\end{figure}

Consider the interaction of an atom (say oxygen atom) with a series of transition metal surfaces.
Within $d$-band model, 
it has been shown that~\cite{nilsson_chemical_2008}
one-electron energy differences do contain information about bonding trends
and surface interaction is assumed to take place in two consecutive steps:
the discrete adsorbate valence state first interact with the $sp$-band of surface, resulting in a broadening of the adsorbate state.
This part of interaction is attractive and believed to vary very little across different transition metal surfaces,
as the coupling matrix element of the adsorbate $sp$ state and the surface $sp$ band is approximately the same for all transition metal surfaces.
The distinguishing part comes from the second step of interaction concerning the more spatially localized $d$ band of surface:
the surface $d$-band hybridizes with the broadened adsorbate state, forming bonding states that are fully occupied, and anti-bonding states (see Figure~\ref{fig:d-band-model} for illustration) 
with extent of filling dependent on the $d$-band center with respect to the Fermi level, i.e.,
\bea 
\varepsilon_d = \int_{-\infty}^{\infty} (\varepsilon - \varepsilon_F) g_d(\varepsilon) \mathrm{d}\varepsilon, 
\eea 
where $\varepsilon_F$ is the energy associated with Fermi level and $g(\varepsilon)$ is the density of state projected to the $d$-orbital, (see subsection \ref{sec:fundamental} for definition) of
the surface atom(s) that are in contact with the adsorbate.
More specifically, when the $d$-band center shifts to a higher position, the filling of anti-bonding states decreases as the anti-bonding states have been pushed above the Fermi level (thus empty), 
meaning the adsorbate-surface system is less destabilized and the resulting bond is stronger.

The formal derivation of the underlying theory of $d$-band model rests upon
the more fundamental Newn-Anderson chemisorption model~\cite{anderson_localized_1961,newns1969,gomer_chemsorption_1975}
and is detailed in the book by Nilsson et al.~\cite{nilsson_chemical_2008}

\section{Theory}
In this section, we summarize relevant theories that are essential to the development of Fermi softness.

\subsection{The fundamentals} \label{sec:fundamental}

Within the Kohn-Sham framework of density functional theory (DFT), the single-electron eigenfunction satisfies:
\bea 
\left [ -\frac{1}{2}\nabla^2 + v(r) \right ]\psi_i(r) = \varepsilon_i \psi_i(r),
\eea 
where $\varepsilon_i$ is the eigenvalue of the eigenstate $ \psi_i (r)$, $v$ is the effective potential experienced by any electron in the system.

The local density of states (LDOS) is defined by
\bea
g(\varepsilon, r) = \sum_i \psi_i^* (r)\psi_i (r) \delta(\varepsilon-\varepsilon_i),
\eea 
where $\delta$ is the Dirac delta function, and 
in practise usually writen as some finite-width normalized Gaussian function.
For periodic system, we have to also include a subscript $k$, indicating the phase factor of basis functions when forming $\psi_{i\fatk}(r)$, i.e.,
\bea \label{eq:ger}
g(\varepsilon, r) = \sum_i \sum_{\fatk} \psi_{i\fatk}^* (r)\psi_{i\fatk} (r) \delta(\varepsilon-\enk),
\eea 
and in this case $i$ labels the band index.
Integrating eqn.~\ref{eq:ger} with respect to spatial coordinates leads to the the total density of states,
\bea \label{eq:ge}
g(\varepsilon) = \int_{-\infty}^{\infty} g(\varepsilon, r)\mathrm{d}r 
\eea 
where $\mathrm{d}r = \mathrm{d}x_1\mathrm{d}x_2\mathrm{d}x_3$ and $-\infty < x_j < \infty,~j=1,2,3$).
While in order to get the electron density of the whole system, integration with respect to energy has to be done, but up to Fermi level only.
This could be automatically done by multiplying the integrant with the Heaviside step function ($\Theta(x)$) when integrating,
\bea 
\rho(r) = \int_{-\infty}^{\ef0} g( \varepsilon, r) \mathrm{d}\varepsilon= \int_{-\infty}^{\infty} \Theta (\varepsilon - \mu) g( \varepsilon, r)\mathrm{d}\varepsilon, 
\eea 
where $\mu$ is the chemical potential (which is equal to the energy of Fermi level $\varepsilon_{\mathrm{F}}$ at absolute zero temperature), the detailed form of $\Theta(x)$ is:
\bea 
\Theta(x) = 
\left\{\begin{matrix}1,~x\le 0
\\ 
0,~x>0
\end{matrix}\right.
\eea 
and its derivative is Dirac $\delta$ function. To yield $N$, the total electron number of the system, one could either integrate $\rho(r)$ with respect to $r$ within the whole space,
\bea \label{eq:rho2N}
N = \int_{-\infty}^{\infty} \rho(r) \mathrm{d}r
\eea 
or through integrating $g(\varepsilon)$ with respect to energy up to $\varepsilon_{\rm F}$:
\bea \label{eq:ge2N}
N = \int_{-\infty}^{\varepsilon_{\rm F}} g(\varepsilon) d\varepsilon.
\eea 

Due to the locality of electronic systems~\cite{kohn_locality_1978,bader_nearsightedness_2008,prodan_nearsightedness_2005, fias2017}, surface atoms are of primary importance in determining the interaction strength between itself and adsorbate. Therefore, we need to project the DOS to surface atom (for instance, atom with index $I$) in contact with adsorbate:
\bea \label{eq:gInlm}
g^{(I)}_{nlm}(\varepsilon) = \sum_i \sum_{\fatk}
| \langle \phi_{nlm}^{(I)}(r) | \psi_{i\fatk}(r) \rangle |^2 \delta(\varepsilon - \varepsilon_{i\fatk})
\eea 
where $nlm$ indicates respectively the principle, angular and magnetic quantum number associated with valence electrons of atom $I$, and $\phi_{nlm}^{(I)}$ represents the corresponding atomic orbital. One could also consider summation of spin quantum number to obtain projected DOS to specific angular channel, e.g., $3d$.

\subsection{Finite electronic temperature density functional theory}
When determining total energy of a periodic system, the band energy (i.e., summation of occupied single-electron state energies) represents one major contribution:
\bea 
E_{\rm band} = \sum_i \frac{1}{\Omega_{\rm BZ}} \int_{\Omega_{\rm BZ}} \enk \Theta (\enk - \mu) \mathrm{d}\fatk
\eea 
where BZ stands for the Brillouin zone.
Restricted by the computational resources, practical calculation of $E_{\rm band}$ has to be approximated by utilising discrete $k$ points (as there are infinite number of such $k$ points in BZ).
That is,
\bea 
E_{\rm band} = \sum_i \sum_{\fatk} w_{\fatk} \enk \Theta(\enk - \mu)
\eea 
where $w_{\fatk}$ is the weight associated with $\fatk$ and the magnitude of $w$'s depend on the k-points sampling scheme.

\begin{figure} 
\includegraphics[scale=0.7]{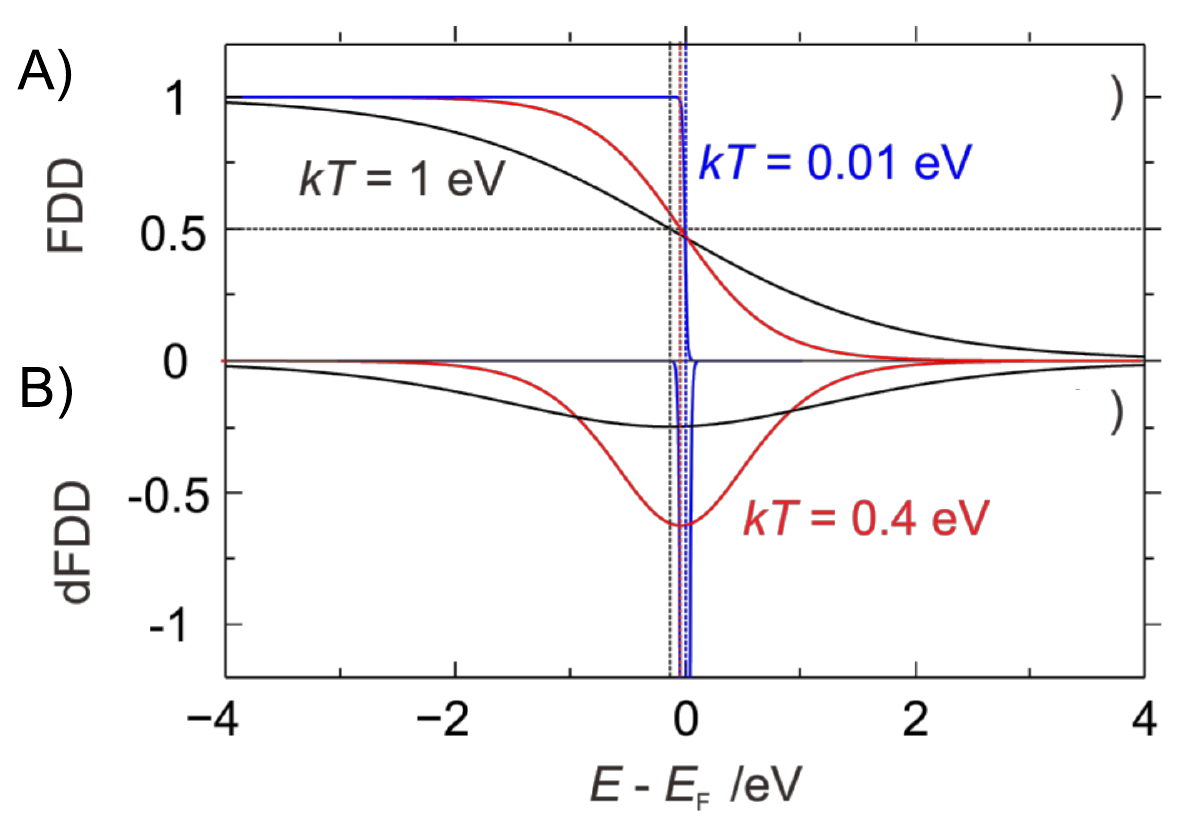}
\caption{Plot of A) Fermi-Dirac distribution (FDD for short) function and B) its derivative (dFDD for short) at three representative electronic temperatures: 0.01 eV, 0.4 eV and 1 eV. One could also see from panel A the approximate equivalence of the chemical potential and the energy associated with Fermi level at these diverse $T_{el}$'s, as indicated by the dashed vertical lines.}
\label{fig:fdd}
\end{figure}

For ground state ($T=0$) gapless systems (e.g., metal and metal surface), the convergence of $E_{\rm band}$ is slow with respect to the number of k points, a ``disaster''
caused by the sudden change of occupation number from 1 to 0 at Fermi level. 
In this case, a large number of k points have to used.
To help reduce the needed k  without lose of integration accuracy, we may resort to partially occupied single electron orbitals in practise.
More specifically, we need to replace the discontinuous unit step function by a smooth one.
Among the multiple choices for this smooth function (e.g., Gaussian function, $N$-th order Methfessel-Paxton (MP) function),
Fermi-Dirac distribution stands out as the physically most meaningful one, i.e.,
\bea \label{eq:fdd}
f(\varepsilon) = \frac{1}{\exp \left( \frac{\varepsilon - \mu}{\sigma} \right) + 1 }
\eea 
where $\mu$ is the chemical potential, $\sigma$ is the parameter controlling the shape of the distribution and $\sigma=kT_{el}$,
with $T_{el}$ being the ``electronic temperature''.
Hereafter, we will use $kT_{el}$ as this temperature instead, with unit eV, for convenience. For visualization of FDD, see Figure~\ref{fig:fdd}.


The replacement of unit step function by a smooth function is usually termed smearing technique and being widely employed in solid state calculations.
However, it has several far-reaching impacts on the system properties.
Most notable is the induced change to energy: the ground state energy is no longer a variational quantity within the exact DFT framework,
and one has to minimize a slightly transformed energy to reach the ground state, i.e., the generalized free energy,
\bea  \label{eq:F}
F = E - \sum_{i} \sum_{k} w_k \sigma S(f_{ik})
\eea 
where $S(f)$ is the electronic entropy of any state with partial occupation $f$, 
\bea  \label{eq:S}
S(f) = -[ f \ln f + (1-f) \ln (1-f) ]
\eea 
The total electronic energy could be obtained by extrapolation to zero electronic temperature and one could verify that $E_0 = E(\sigma \rightarrow 0)=1/2(F+E)$.
Together with Janak's theory~\cite{janak_proof_1978}, and the constraint the total number of electrons being summation of $f_i$ (i.e., $N=\sum_i f_i$), one can verify the consistency among equations \ref{eq:fdd}, \ref{eq:F} and \ref{eq:S} by the method of Lagrange's multiplier.

The impact of smearing on other properties, such as the (local) density of state, is assumed to be small, as defaulted in almost all solid state DFT programs.
Now the expression for LDOS, has to be adapted only for the finite number of $k$ points, i.e.,
\bea 
g(\varepsilon, r) = \sum_i \sum_{\fatk} w_{\fatk} \psi^*_{i{\fatk}} (r) \psi_{i{\fatk}} (r) \delta(\varepsilon-\enk),
\eea 
and eqn.~\ref{eq:ge} still holds true. For properties regarding electron number (i.e., electron density and total electron number of the whole system), 
however, smearing has to be considered explicitly. For electron density, we have
\bea 
\rho(r) = \int_{-\infty}^{\infty} f(\varepsilon) g( \varepsilon, r)\mathrm{d}\varepsilon, 
\eea 
and eqn.~\ref{eq:rho2N} retains its current form regardless. When expressing $N$ as function of $g(\varepsilon)$, eqn.~\ref{eq:ge2N} needs to be also revised, i.e.,
\bea \label{eq:ge2N_Tel}
N = \int_{-\infty}^{\infty} g(\varepsilon) f(\varepsilon) d\varepsilon.
\eea

\subsection{Fermi softness}

\subsubsection{Basic idea} \label{sec:frontier}

As has already been widely acknowledged~\cite{hoffmann_chemical_1988,hammer_why_1995}, both occupied and unoccupied states of solid surface are active when interacting with the adsorbate, though each state (be it occupied or not) may interact differently with the HOMO and LUMO of the adsorbate,
and accordingly, each results in different contribution to the surface bonding.

\begin{figure}[!ht] 
\includegraphics[scale=0.8]{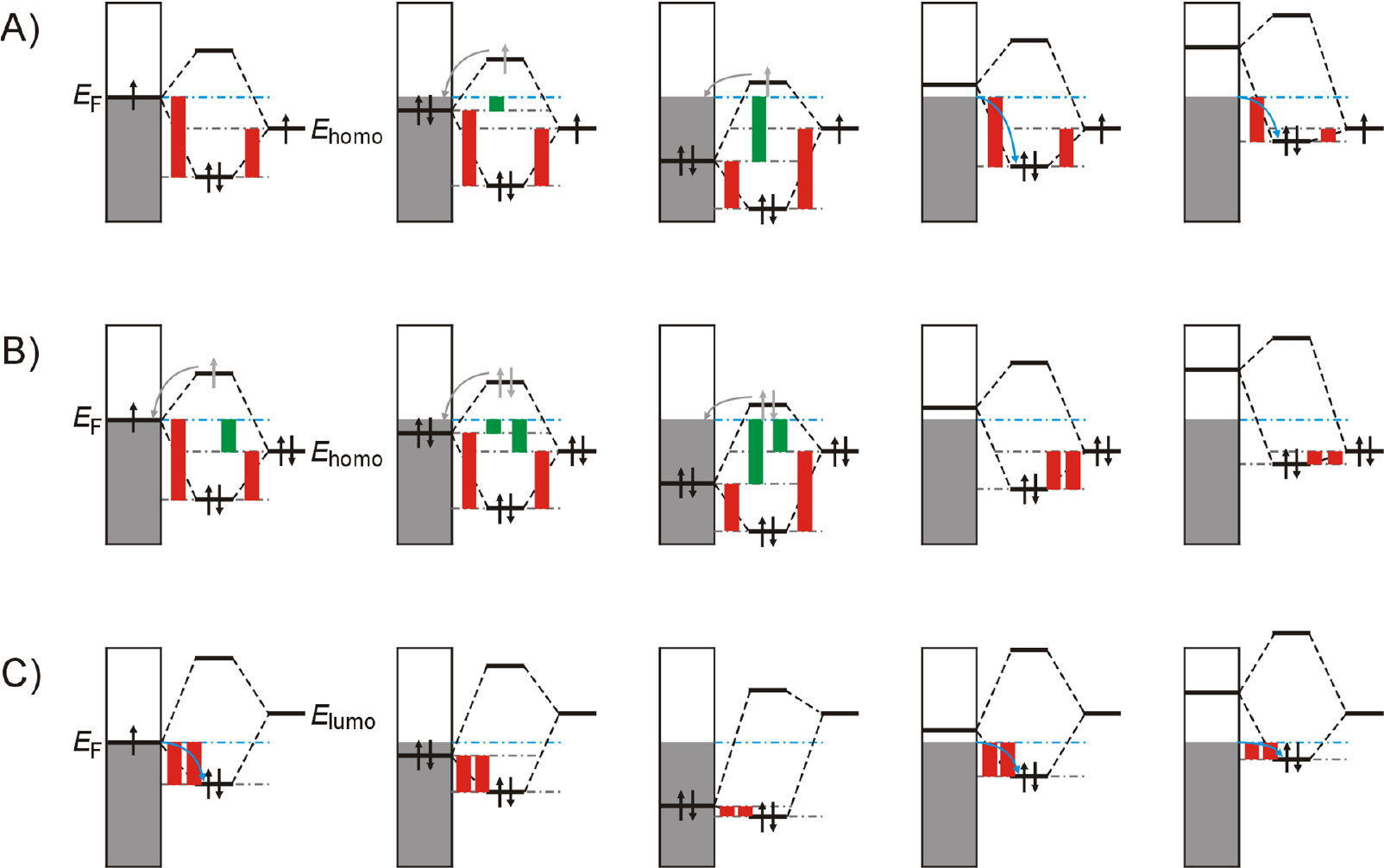}
\caption{Orbital interactions of surface band and three types of adsorbate orbital:  half-occupied HOMO (A), doubly occupied HOMO (B) and empty LUMO (C). The five columns respectively correspond to five different states of surface band, being at Fermi level, slightly below $\varepsilon_{\rm{F}}$, well below $\varepsilon_{\rm{F}}$, slightly above $\varepsilon_{\rm{F}}$ and well above $\varepsilon_{\rm{F}}$. Red bar: magnitude of energy gain after orbital interaction; Green bar: magnitude of energy penalty upon orbital interaction; Solid upward (downward) pointing arrow: electron with spin up (down) occupying some electronic state; Gray arrow: electron occupying some ``intermediate'' electronic state that is about to fill the unoccupied state at the Fermi level of surface band. }
\label{fig:orbint}
\end{figure}

Here, we analyze in detail the energetic consequences for overall fifteen types of orbital interactions, being combinations of three different types of adsorbate electronic states (half-occupied HOMO, fully occupied HOMO and empty LUMO) with five representative electronic states of surface band with varied position relative to Fermi level, as depicted in Figure~\ref{fig:orbint}.

Let the active state of the adsorbate be half-occupied HOMO.
Consider its interaction with the half-filled Fermi level of the surface, there will be energy gain upon bonding for both of the two involved electrons, as they both descend into a lower-energy bonding state.
By shifting the surface electronic state to a state slightly below Fermi level, which is now doubly occupied, the outcome would be slightly different, due to the occurrence of a net ``transfer'' of one surface electron from below Fermi level to Fermi level, causing a small magnitude of energy penalty (i.e., raising of orbital energy of single electron state).
As the energy penalty is small, the total single electron energy (that is, the sum of red bars minus the sum of green bars (if any) in Figure~\ref{fig:orbint}) decreases, i.e., the whole system is stabilized. 
As the surface electronic state moves further away from Fermi level, its interaction with the HOMO of the adsortate becomes weaker, resulting from the larger magnitude of energy penalty, as indicated by the longer green bar in the third column of Fig.~\ref{fig:orbint}A.
Similar weakening happens when we shift the surface electronic state upward from Fermi level (which is now empty), as illustrated in the fourth column of Fig.~\ref{fig:orbint}A.
The major differences (cf. column 2 and 3 in Fig.~\ref{fig:orbint}A) are two-fold: i) no energy penalty happens;
ii) the internal transfer of electrons of surface persists, but with reversed direction, i.e., there is a net transfer of electron from higher energy levels to lower ones within the surface band, further stabilizing the system.

When the HOMO of the adsorbate is fully occupied, similar statements could be made as in the case of singly occupied HOMO.
While notable differences would be expected when the active orbital of the adsorbate is the empty LUMO.
As displayed in Fig.~\ref{fig:orbint}C, all five types of interaction tend to stabilize the system, free of energy penalty, differing primarily in the magnitude of energy gain.
Of particular interest are the last two types of interaction in Fig.~\ref{fig:orbint}C, where the interaction is between non-occupied state, yet there is still energy gain. This is unique to surface interaction, as was proposed decades ago by Hoffman~\cite{hoffmann_how_1987}.



Based on the analysis above, we could draw one revealing conclusion:
the closer the state is to the Fermi level, the greater its contribution to bonding.
Hence, to qualitatively correct describe the reactivity of a surface, one may propose a quantity as a function of the density of states ($g(\varepsilon)$) and some weight function ($w(\varepsilon)$) that peaking at the Fermi level.
The sum of all the weighted contributions ($\int g(\varepsilon)w(\varepsilon)\mathrm{d}\varepsilon$) is conceived to act as a reactivity descriptor of a surface.

Now the problem boils down to what form of weight function $w(\varepsilon)$ should one choose.
Among the many options, 
herein, we assign $w(\varepsilon)$ to the derivative of the Fermi–Dirac distribution (dFDD for short) function at a non‐zero electronic temperature, i.e., 
\bea 
w(\varepsilon) = -f'(\varepsilon) = -\frac{d f(\varepsilon)}{d\varepsilon} = \frac{ \exp \left ( \frac{\varepsilon - \mu}{\sigma} \right ) }{ \left [ \exp \left( \frac{\varepsilon - \mu}{\sigma} \right) + 1 \right]^2 },
\eea 
which peaks at $E_F$ as required
and diminishes to zero as single electron state moves away from the Fermi level.
The resulting weighted sum of the reactivity contribution is dubbed Fermi softness (labeled as $s_F$) for reasons to be elaborated shortly and can be expressed as
\bea 
s_F (\varepsilon, r) = -\sum_i \sum_{\fatk} w_{\fatk} f'(\varepsilon) \psi_{i\fatk}^* (r) \psi_{i\fatk} (r) \delta(\varepsilon-\varepsilon_{i\fatk}),
\eea 
where the spreading of $f'(\varepsilon)$ can be changed by adjusting the parameter electronic temperature $\sigma=kT_{el}$ (see eqn.~\ref{eq:fdd}).
Similar to how we obtain the electron number $N$ and $\rho(r)$, the global softness $S_F$ and its local version $s_F(r)$ have the following expressions:
\bea 
s_F(r) &=& \int_{-\infty}^{\infty} s_F(\varepsilon, r) \mathrm{d}\varepsilon \\ 
S_F &=& \int_{-\infty}^{\infty} s_F(r) \mathrm{d}r
\eea 
When integrating $s_F(\varepsilon, r)$ w.r.t. $r$ alone, we end up with a quantity $s_F(\varepsilon)$ similar to the density of states, i.e.,
\bea 
s_F(\varepsilon) = \int_{-\infty}^{\infty}  s_F(\varepsilon, r) \mathrm{d}r
\eea 
Further integration of $s_F(\varepsilon)$ w.r.t. energy also gives the global softness $S_F$,
\bea 
S_F = \int_{-\infty}^{\infty} s_F(\varepsilon) \mathrm{d}\varepsilon
\eea

We note by passing that dFFD is normalized, regardless of the value $kT$ takes, i.e.,
\bea 
- \int_{-\infty}^{\infty} f'(\varepsilon) = 1
\eea 
This feature is coveted as it is the density of states that enters as the sole factor determining the reactivity of a surface.

\begin{figure} 
\includegraphics[scale=1.3]{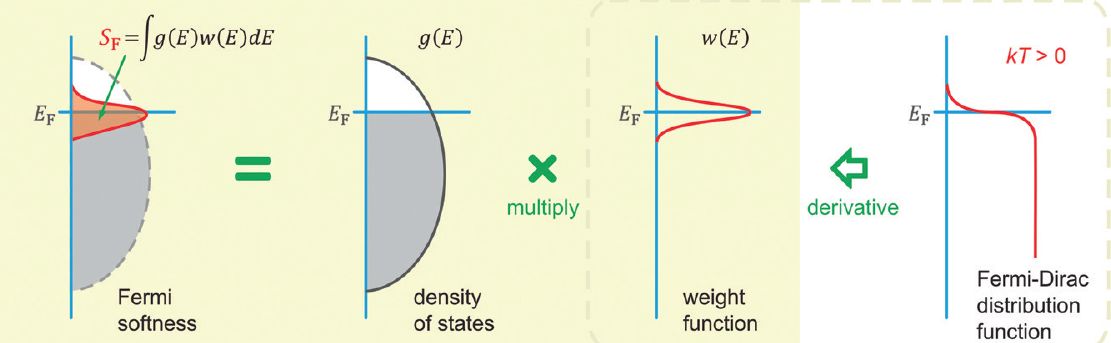}
\caption{Graphical illustration of $S_F$, defined as
a weighted sum of the density of states $\int g(\varepsilon)w(\varepsilon)\mathrm{d}\varepsilon$, where the weight function $w(\varepsilon)$ is chosen as the derivative of the Fermi–Dirac distribution function at non-zero electronic temperature. 
}
\label{fig:gsf}
\end{figure}

Due to the existence of locality of electronic systems aforementioned, we have to extract the contribution to $S_F$ from some surface atom (say atom with index $I$) only. There exists several approaches to do this. One straightforward way is to partition the space into atomic contributions (e.g., Bader scheme, Voronoi scheme, or simply Wigner-Seitz scheme, etc.) and then integrate $s_F(r)$ within the subspace associated with atom $I$, i.e.,
\bea 
S_F^{(I)} = \int_{\Omega_I} s_F(r)\mathrm{d}r
\eea 
where $\Omega_I$ is the space partitioned to atom $I$.
Due to the fact that space partitioning could be time-consuming (especially true for Bader scheme for large systems), an alternative $s_F(\varepsilon)$ based approach may be adopted, in which we have to obtain the density of states projected to surface atom $I$ first, and then employ the following equation to calculate $S_F^{(I)}$ (see Fig.~\ref{fig:gsf} for a graphical illustration):
\bea
S_F^{(I)} &=& \int_{-\infty}^{\infty} s_F^{(I)}(\varepsilon) \mathrm{d}\varepsilon \\ 
&=&  \int_{-\infty}^{\infty} g^{(I)}(\varepsilon) f'(\varepsilon) \mathrm{d}\varepsilon
\eea 
where $s_F^{(I)}(\varepsilon)$ is the projected energy-resolved Fermi softness of surface atom $I$, and
$g^{(I)}(\varepsilon)$ is the total density of states projected to $I$, including all possible angular components of atom $I$, i.e.,
\bea 
g^{(I)}(\varepsilon) = \sum_{nlm\in I} g^{(I)}_{nlm}
\eea 
where $g_{nlm}^{(I)}(\varepsilon)$ is defined in eqn.~\ref{eq:gInlm}.

Before leaving this subsection, it is necessary to clarify the exact meaning of being global or local.
Here, global refers to a single scalar quantity, for instance, electron number, or the global Fermi softness ($S_F$).
Note that $S_F$ may also refers to a specific surface atom, i.e., $S_F^{(I)}$, and it is the default meaning of $S_F$ whenever we mention $S_F$ hereafter, unless otherwise stated.
When speaking of local picture, we mean there is a scalar value associated with every point in the three dimensional space, for instance, charge density and the local Fermi softness $s_F(r)$.

\subsubsection{First-principle view of Fermi softness}

$S_F$ and $s_F(r)$ are not just some random quantities that happen to serve well as reactivity indices (as will be shown shortly), rather, they are meaningful and rooted in DFT. Here, we derive in detail how they naturally arise within conceptual DFT.

\begin{figure} 
\includegraphics[scale=0.7]{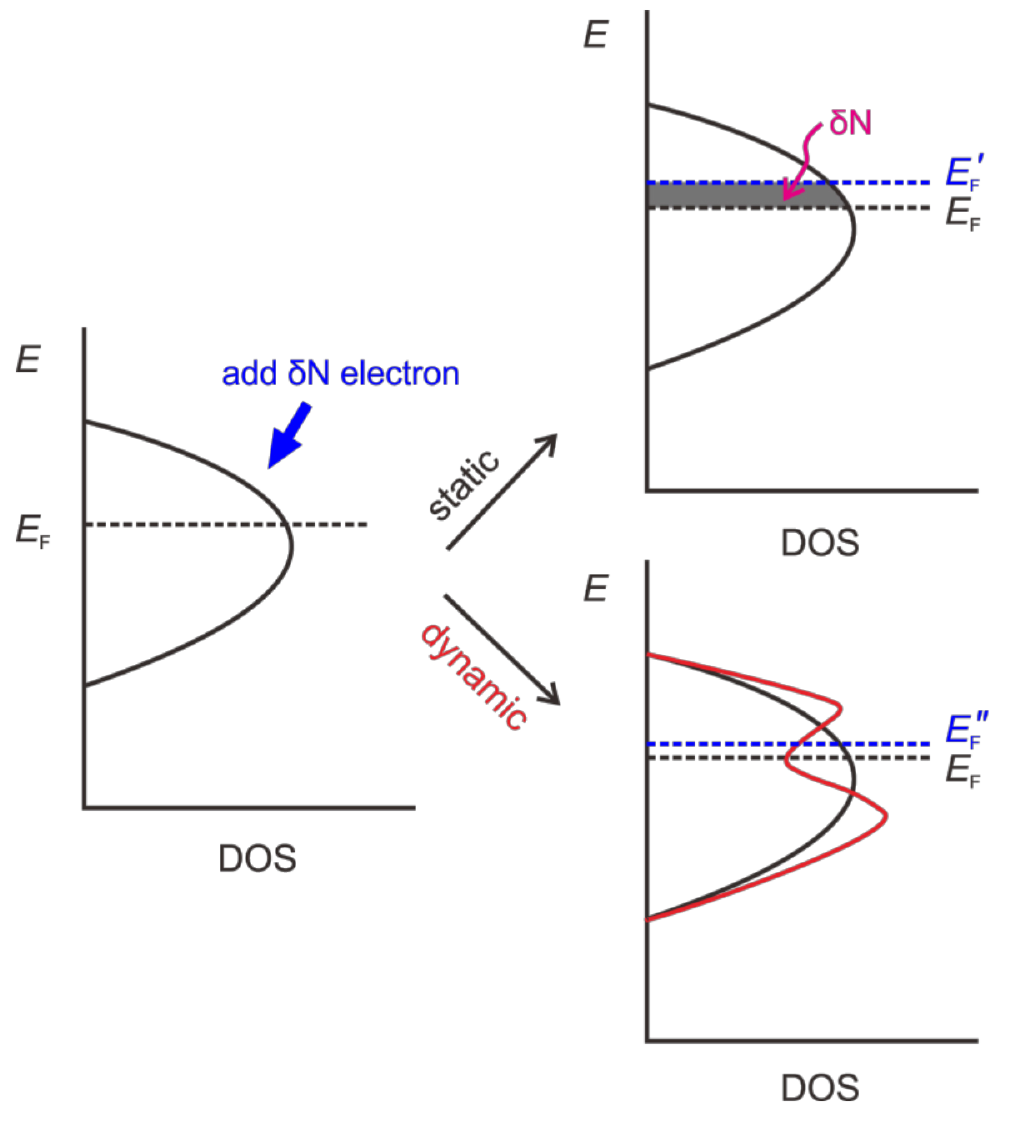}
\caption{Illustration of static and dynamic adaptation of DOS under external perturbation by adding a small amount of electron $dN$. The effect of perturbing $\mu$ by a small amout d$\mu$ could be understood in a similar fashion.
Note that the dynamic adaptation is closer to reality.}
\label{fig:dgedn}
\end{figure}

We start from the classic concept global ``softness”, but within the finite electronic temperature framework: 
\bea 
S = \left (  \frac{\partial N}{\partial \mu} \right )_{T_{el}, v}
\eea 
where $v$ indicates external potential.
Plugging in eqn.~\ref{eq:ge2N_Tel} and complete the square, we have
\bea \label{eq:dNdu}
S = \int g(\varepsilon) \frac{\partial f(\varepsilon )}{\partial \mu} d\varepsilon +  \int \frac{\partial g(\varepsilon)}{\partial \mu} f(\varepsilon ) d\varepsilon
\eea 
As the equation above implies, a infinitesimal change in $\mu$ induces not mere change in $N$, but also accompanied by change in density of states (and more generally electronic structure, see Fig~\ref{fig:dgedn} for an intuitive graphical illustration), though which is likely to be relatively smaller in magnitude compared to the first term in the right hand side (RHS) of eqn.~\ref{eq:dNdu}.

Consequently, the second term in the RHS of eqn.~\ref{eq:dNdu} is often neglected. Note that in some cases, the second term may not be negligible and its contribution must be assessed carefully (to be elaborated later).

Assume the induced change in electronic structure is very small for now, and
\bea 
\frac{d f(\varepsilon)}{d\mu} = \frac{ \frac{1}{\sigma} \exp \left ( \frac{\varepsilon - \mu}{\sigma} \right ) }{ \left [ \exp \left( \frac{\varepsilon - \mu}{\sigma} \right) + 1 \right]^2} = -\frac{d f(\varepsilon)}{d\varepsilon}  
\eea

As within a wide range of $T_{el}$, $\mu$ is approximately equal to the energy corresponding to Fermi level (\textcolor{red}{see Fig.~\ref{fig:fdd} for illustration}), therefore, eqn.~\ref{eq:dNdu} could be simplified as
\bea \label{eq:sf0}
S_F \approx -\int_{-\infty}^{\infty} g(\varepsilon) f'(\varepsilon) d\varepsilon
\eea 
where $f'(\varepsilon) = \frac{\partial f(\varepsilon)}{\partial \varepsilon}$.

Similarly, the local version of $S_F$ is $s_F(r)$,
\bea  \label{eq:sfr}
s_F(r) &=& \left ( \frac{\partial \rho(r)}{\partial \mu} \right)_{T_{el}, v} \\ 
&=& \int_{-\infty}^{\infty} g(\varepsilon,r) \frac{\partial f(\varepsilon )}{\partial \mu} d\varepsilon +  \int_{-\infty}^{\infty} \frac{\partial g(\varepsilon,r)}{\partial \mu} f(\varepsilon ) d\varepsilon \label{eq:sf0Term2}
\eea 
Neglecting the second term in the RHS of the equation above, we have,
\bea \label{eq:sf}
s_F(r) \approx -\int_{-\infty}^{\infty} g(\varepsilon,r) f'(\varepsilon) d\varepsilon
\eea
Combining eqn.~\ref{eq:sf0} and~\ref{eq:sfr}, Fukui function at finite electronic temperature could be written as
\bea 
f_T(r) &=& \left ( \frac{\partial \rho(r)}{\partial N} \right)_{T_{el}, v} \\
&=& \left ( \frac{\partial \rho(r)}{\partial \mu} \right)_{T_{el}, v} \left ( \frac{\partial \mu}{\partial N} \right)_{T_{el}, v} \\ 
&=& \frac{s_F(r)}{S_F} \label{eq:fT1}\\ 
&\approx& \frac{\int_{-\infty}^{\infty}  g(\varepsilon, r)f'(\varepsilon)d\varepsilon}{\int_{-\infty}^{\infty} g(\varepsilon)f'(\varepsilon) d\varepsilon} \label{eq:fT2}
\eea 
where the subscript $T$ of $f$ indicates finite electronic temperature, so as to be distinguished from Fukui function ($f(r)$) at absolute zero temperature.
Therefore, $s_F(r)$ can be also written as
\bea \label{eq:sf2}
s_F(r) \approx S_F f_T(r)
\eea 
which is likely to be a more accurate expression of $s_F(r)$ (cf. eqn.~\ref{eq:sf}) as $g(\varepsilon, r)$ is relatively more sensitive to perturbation in $\mu$, while the integrated form of $g(\varepsilon, r)$ over space would be relatively more robust against external perturbation as the induced spatial changes, to some degree, would be integrated out.
More specifically, we could rearrange terms in equations \ref{eq:fT1} and \ref{eq:fT2} as
\bea \label{eq:sfTerm2} 
s_F(r) &=& f_T(r) \left (\int_{-\infty}^{\infty}  \mathrm{d}r  \int_{-\infty}^{\infty} g(\varepsilon,r) \frac{\partial f(\varepsilon )}{\partial \mu} d\varepsilon + \int_{-\infty}^{\infty} \mathrm{d}r \int_{-\infty}^{\infty} \frac{\partial g(\varepsilon,r)}{\partial \mu} f(\varepsilon ) d\varepsilon \right)\\ 
&=& f_T(r) \int_{-\infty}^{\infty} g(\varepsilon) \frac{\partial f(\varepsilon )}{\partial \mu} d\varepsilon + f_T(r) \int_{-\infty}^{\infty} \frac{\partial g(\varepsilon)}{\partial \mu} f(\varepsilon ) d\varepsilon,
\eea 
The second term in the RHS of eqn. \ref{eq:sfTerm2} would be smaller in magnitude compared to the second term in eqn. \ref{eq:sf0Term2}.
Nevertheless, in practise we would stick to eqn. \ref{eq:sf} for computation of $s_F(r)$, due to the challenge posed by dealing with fractional charges for solid surface within the finite difference approach to the computation of $f_T(r)$ in eqn. \ref{eq:sf2}.

Integrating $s_F(r)$ w.r.t. $r$ within the whole space, we obtain the global Fermi softness $N_F$,
\bea 
S_F = \int_{-\infty}^{\infty} s_F(r)\mathrm{d}r 
\eea 
As in the special case of $T_{el}=0$, FDD degenerates into Heaviside step function $\Theta(x)$ and the derivative of FDD is simply Dirac delta function, i.e., 
\bea 
-f’(\varepsilon-\mu) = \delta(\varepsilon-\mu)
\eea 
Accordingly,
\bea 
f_T(r) &\approx& \frac{\int_{-\infty}^{\infty}  g(\varepsilon, r) \delta(\varepsilon-\mu)d\varepsilon}{\int_{-\infty}^{\infty} g(\varepsilon)\delta(\varepsilon-\mu) d\varepsilon} \\ 
&=& \frac{g(\ef0,r)}{g(\ef0)}
\eea 
where we have used $\ef0 = \mu$ when $T_{el}=0$. The thus-obtained expression of $f_{T_{el}=0}$ is exactly what Yang and Parr once derived~\cite{yang_hardness_1985}. This indirectly verifies all relevant equations above for finite temperature.

\begin{figure} 
\includegraphics[scale=1.2]{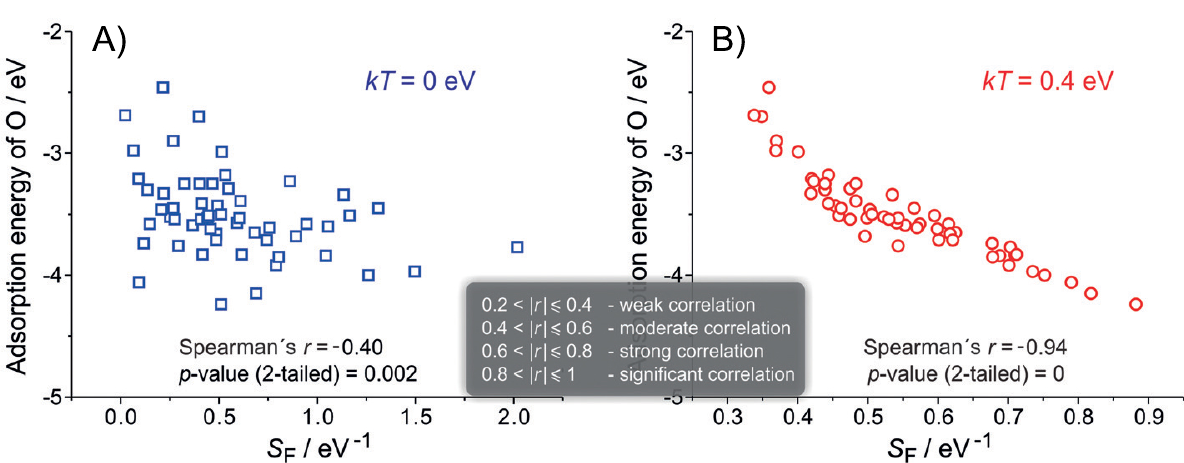}
\caption{Comparison of correlations between oxygen adsorption energy ($\Delta E_{O}$) on Pt-related surfaces and $S_F$ of surface Pt atom at two electronic temperatures: 0 eV (A, no spreading in dFDD) and 0.4 eV (B). How well $\Delta E_{O}$ is correlated with $S_F$ is measured by the Spearman's $r$ and larger absolute value indicates better correlation.}
\label{fig:corr}
\end{figure}

To verify that finite electronic temperature does play a pivotal role in determining the reactivity of a surface,
we have examined the correlation between oxygen adsorption energy on Pt-ML/M surface (where M is transition metal substrate),
at two distinct electronic temperatures: 0 eV and 0.4 eV.
In the former case, the derivative of FDD degenerates into a single Heaviside step function and $S_F$ is precisely the density of states at Fermi level ($g(\varepsilon_F)$).
As displayed in Fig.~\ref{fig:corr}, $S_F$ at 0 eV shows a rather weak correlation with the oxygen adsorption strength, as is always suggested by the small Spearman's $r$ value (-0.40) in absolute value.
In contrast, the correlation between oxygen adsorption energy and $S_F$ becomes pronounced at a finite electronic temperature of 0.4 eV, i.e., Spearman's $r$ is as great as 0.94 in absolute value.
When $kT_{el}$ is set to other values, better correlations are also observed (cf. the case of zero $kT_{el}$), but the best correlation is achieved at $kT_{el}\sim 0.4$ eV.

\subsubsection{Fermi softness vs d-band center}

There are three major differences between d-band model and Fermi softness: i) Fermi softness offers both global and spatial picture of surface reactivity, while d-band model offers only the global picture.
ii) Fermi softness describes the response of the whole electron density with respect to change in chemical potential, i.e., it includes the contribution from both s-, p- and d-orbitals, 
while d-band model, as its name implies, considers only projection to d-orbitals (the projection amplitude does depend on projection amplitude to the other angular channels though).
Note that this does not suggest better performance of Fermi softness model (cf. d-band model), as the contribution to interaction from sp-band is similar across different transition metal surfaces.
By subtracting this part of contribution, the relative magnitude of Fermi softness across transition metal surfaces would not change. 
That is, at least for transition metal surface, sp-band has minor effect on the relative strengths of Fermi softness.
However, for other gapless systems, such as sp-block metal surface or non-metal surfaces, Fermi softness may stand out as a more competitive descriptor due to its response nature.
iii) d-band model is theoretically more rigorous than Fermi softness, as the former offers a direct link between chemisorption strength and d-band center (though it is much approximated) based on simplified quantum model. 
Fermi softness, like most CDFT-based descriptor, is lack of such direct relationship and often assumed to describe only the early stages of interaction.

Regarding the first point above, it is worthwhile to mention that one could also propose a quantity similar to $s_F(r)$ for d orbitals, i.e.,
\bea 
g^{(I)}_{d}(r) = 
\sum_m \sum_i \sum_{\fatk} w_{\fatk}
 \phi_{m}^{(I)*}(r) \psi_{i\fatk}(r)  
\eea 
where $m$ indicates the magnetic quantum number of valence $d$ electrons of atom $I$. One could also further sum up contributions from all surface atoms.
This is, however, not quite meaningful, as suggested by the frontier band view on interaction elaborated in section \ref{sec:frontier} and hardly could one establish any connection to a more fundamental theory.

In principle, both d-band model and Fermi softness are applicable to systems sharing some similarity only, but for somewhat different reasons.
More specifically, for Fermi softness, the interaction types/trends in early stages are similar, so are the interaction types/trends in late stages (which determine the final interaction strength) and therefore the relative interaction strengths persist throughout.
For d-band model, similar coupling matrix element between adsorbate state and surface $d$ band would be desired~\cite{nilsson_chemical_2008}.

\section{Applications}
In this section, we demonstrate the usefulness of Fermi softness, including both its global and local picture through a few representative applications. For more use cases, the reader is referred to the original publication~\cite{huang_spatially_2016}.

\subsection{Correlation with adsorption energy}

In Figure \ref{fig:corr}, we have already demonstrated that the global Fermi softness as a reactivity index exhibits a nice correlation with oxygen adsorption energy for Pt-derived metal surfaces.
Here we offer two more examples.

\begin{figure}[ht!] 
\includegraphics[scale=0.9]{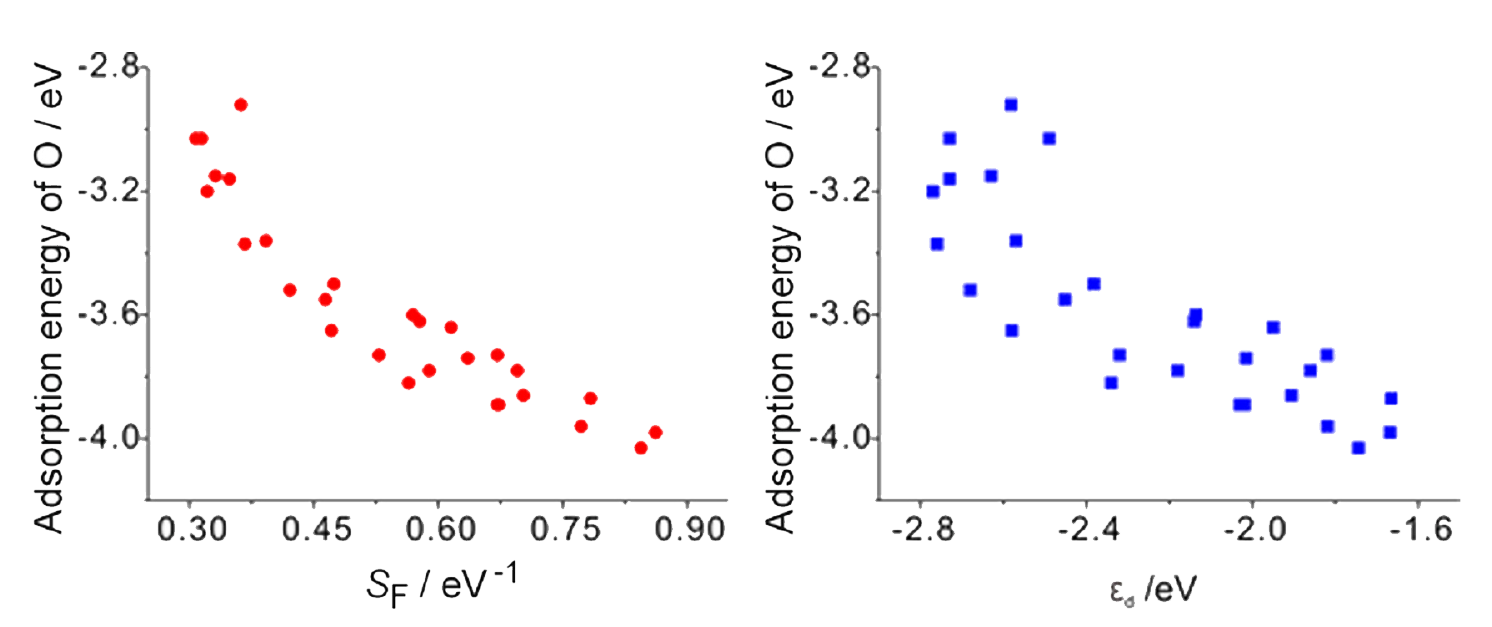}
\caption{Correlation between the global Fermi softness ($S_F$) and the surface reactivity as indicated by the oxygen adsorption strength on Pd-derived surfaces (see the supplementary material of reference~\cite{huang_spatially_2016} for details). For comparison, the correlation of the d-band center for the same systems is also shown.}
\label{fig:sf_vs_ed}
\end{figure}

The first one is for Pd(111) and supported Pd monolayer on various transition metal surfaces.
The property investigated is still oxygen adsorption energy.
Again, an approximate monotonic trend is observed: as the global Fermi softness of surface Pd atom increases, the surface reactivity is enhanced, resulting in stronger chemical bond.
As a comparison, correlation of oxygen adsorption energy and d-band center for the same set of systems, is also plotted, as displayed in the right panel of  Figure \ref{fig:sf_vs_ed}.
The latter correlation to $\varepsilon_d$ is more scattered, and notably worse than the former correlation with $S_F$.
The same statement could be made for Pt and other transition metal derived systems.
This seems to suggest that the global Fermi softness is a very robust surface reactivity descriptor.

\begin{figure}[ht!] 
\includegraphics[scale=0.9]{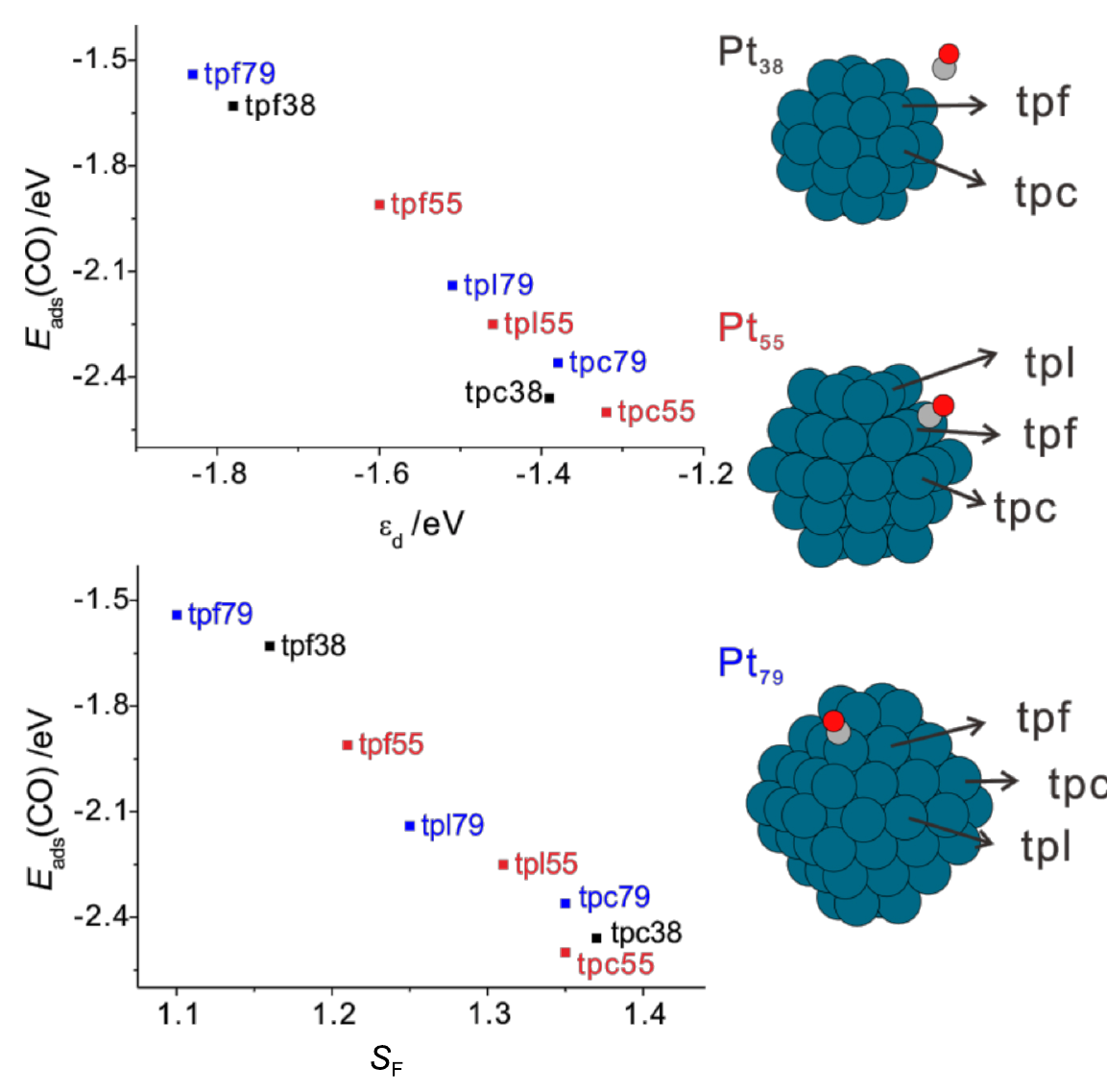}
\caption{Demonstration of the correlation between the Fermi softness ($S_F$) and the reactivity of various top sites (tpf, tpc and tpl, as shown in the inset) on Pt nanoparticles as represented by the CO adsorption strength (upper panel). For comparison, correlation with the d-band center is also presented (lower panel). tpl79: top site on Pt atom in Pt nanoparticle made up of 79 Pt atoms.}
\label{fig:corr_PtNP}
\end{figure}

To further support our finding, more numerical results are necessary.
As such, we consider further the adsorption of a saturated molecule carbon monoxide (CO) on a series of Pd, Pt and Au nanoparticles (NP), with results for Pt NPs shown in Figure \ref{fig:corr_PtNP}.
The overall correlation of $\Delta E_{ads}^{\mathrm{CO}}$ with $S_F$ is very good, and resembles much the correlation with d-band center.
Similar correlations could be found for other NPs (not shown here).

\subsection{Active sites of MoS$_2$}
The local picture of Fermi softness is particularly valuable when dealing with heterogeneous surface structure,
which represents a wider range of catalysts in real-world applications.
Here, we demonstrate its power through only one example: a one-dimensional (1D) MoS2 edge, which is responsible for the high catalytic activity towards hydrogen evolution reaction~\cite{tsai_tuning_2014,bollinger_mos2_2001,jaramillo_identification_2007}.

\begin{figure} 
\includegraphics{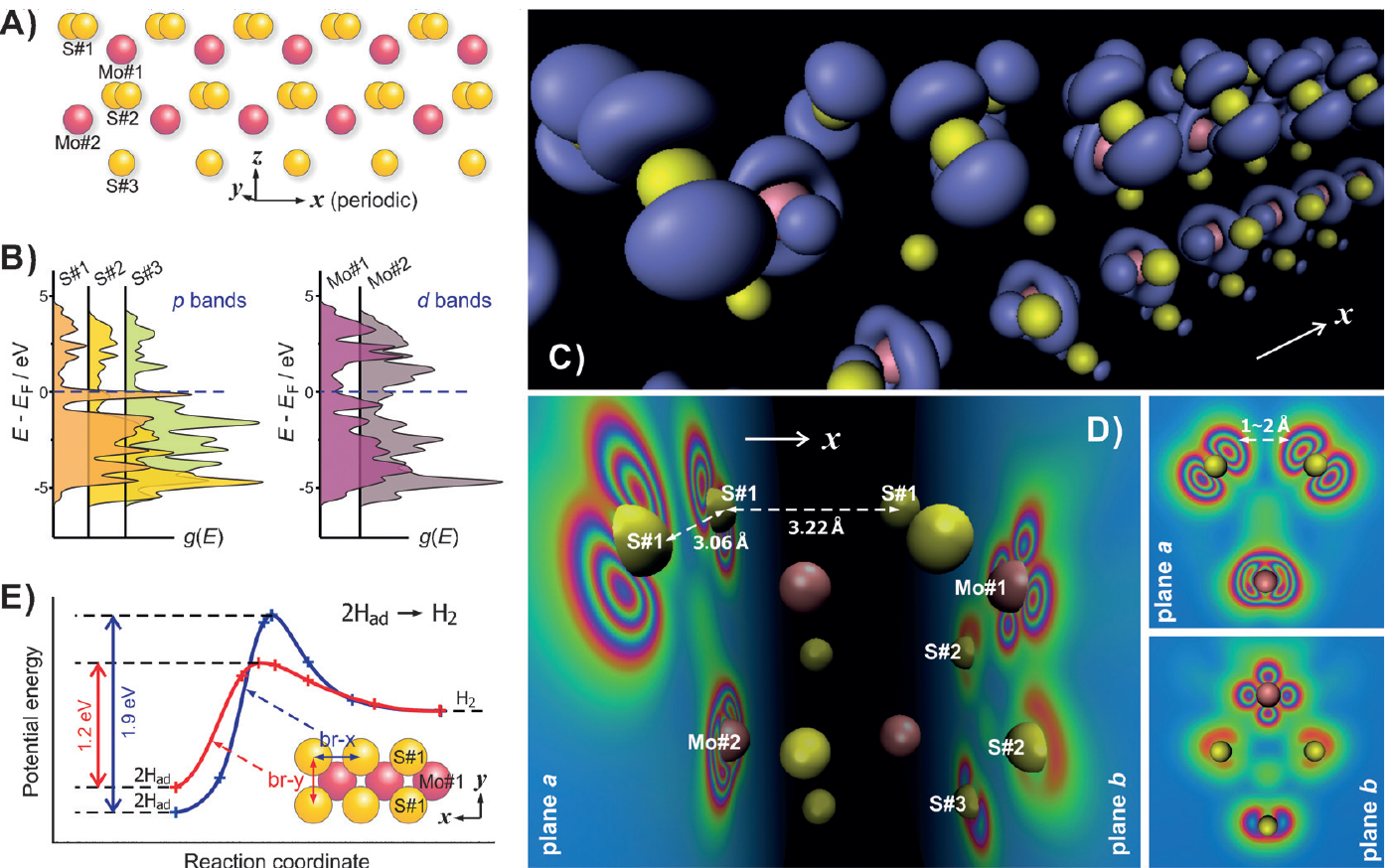}
\caption{Spatially resolved chemical reactivity of a 1D MoS$_2$ edge. A) The geometrical structure of MoS$_2$, with three typical local sulfur environments (S\#1, S\#2, and S\#3) and two typical molybdenum atom environments (Mo\#1 and Mo\#2). B) The projected density of states ($g(\varepsilon)$) of specific sulfur and molybdenum atoms. C) A 3D perspective of the local Fermi softness $s_F(r)$, as represented by some isosurface (blue). D) The $s_F(r)$ is plotted as projection onto two planes normal to the
x direction: plane \textbf{a} intersects two S\#1 and one Mo\#2 atoms, and plane \textbf{b} contains one Mo\#1, two S\#2, and one S\#3 atoms. E) Reaction energy profiles for the reaction 2H$_{\mathrm{ad}}\rightarrow $H$_2$ at two kinds of bridge sites (br-x or br-y, see the inset) on the S\#1 dimer edge, obtained from nudged elastic band calculations. }
\label{fig:mos2}
\end{figure}

The structure of 1D MoS$_2$ is displayed in Figure \ref{fig:mos2}A and exhibits much richer local atomic environments than previously studied transition metal derived surfaces.
More specifically, it covers three types of sulfer atoms, including S\#1, S\#2 and S\#3, as well as two types of Mo atoms, i.e., Mo\#1 and Mo\#2.
Among all the S sites, the dimer consisting of two S\#1 atoms (along y direction, see Figure \ref{fig:mos2}A) stands out as the most prominent local environments, i.e., around which the reactivity distribution of $s_F(r)$ exhibits
a magnitude significantly larger than that of any of the rest S environments,
which is totally consistent with the experimental finding~\cite{bollinger_mos2_2001,jaramillo_identification_2007}.
The sharp peak around Fermi level in the projected DOS plot in Figure \ref{fig:mos2}B also support the observation.
More intriguing is the tilted p-orbital-like distribution of $s_F(r)$ around S\#1, which reveals the subtle information that the reactivity of the S\#1 dimer edge is spatially anisotropic:
the reactivity field over the aforementioned S dimer is stronger than between the two S\#1 atoms lying along x direction.
In particular, the distance between the two centers of the crown of $s_F(r)$ of two S\#1 atoms in the S dimer matches the inter-atomic distance of two H atoms in H$_2$ and is therefore ideal for catalyzing the formation reaction of H$_2$ from two adsorbed H atoms.
To verify such a hypothesis, we have calculated the reaction profile of H-H bond formation reaction 2H$_{\mathrm{ad}}\rightarrow $H$_2$ at two kinds of bridge sites:
br-x (intra-dimer site) or br-y (inter-dimer site, see the inset of Figure \ref{fig:mos2}E) on the S\#1 dimer edge.
The computational results clearly show that the barrier at the br-x site is significantly lower compared to that at the br-y site,
regardless of the direction of the reaction.
This is in full agreement with the anisotropic nature suggested by $s_F(r)$, 
an insight previously unknown and could not be unraveled by any other reactivity descriptors, to the best of our knowledge.

\section{Conclusion and Outlook}
To recap, we have reviewed the detailed theory of Fermi softness, and some representative applications.
Numerical evidences suggest that Fermi softness could serve as a very useful reactivity descriptor for solid surface due to its robustness as well as low computational cost.
While the d-band model is limited to transition metal surfaces and nanoparticles,
the potential applicability of Fermi softness is largely unlimited,
thanks to its response nature,
as well as its deep connection to the conceptual density functional theory.
Currently, only a small amount of applications have been reported,
its power to tackle more complicated systems and ultimately help with catalysts design still awaits to be fully unleashed in the future.


\section*{Acknowledgement}
B.H. acknowledges funding from the Swiss National Science Foundation (No. PP00P2\_138932 and 407540\_167186 NFP 75 Big Data). This research was partly supported by NCCR MARVEL, funded by the Swiss National Science Foundation.

\bibliographystyle{ieeetr} 
\bibliography{sf}%

\end{document}